\documentstyle[11pt,emulateapj]{article}
\slugcomment{To appear in {\it The Astronomical Journal}}

\newcommand{\etal}{{\it et\thinspace al.}\ }
\newcommand{\kms}{km\thinspace s$^{-1}$}
\newcommand{\simlt}{\ {\raise-.5ex\hbox{$\buildrel<\over\sim$}}\ }

\lefthead{Salzer \etal}
\righthead{KPNO International Spectroscopic Survey}

\begin{document}

\title{The KPNO International Spectroscopic Survey.  \\II. H$\alpha$-selected Survey List 1.}

\author{John J. Salzer\altaffilmark{1,2} and Caryl Gronwall\altaffilmark{1,3}}
\affil{Astronomy Department, Wesleyan University, Middletown, CT 06459; 
slaz@astro.wesleyan.edu}

\author{Valentin A. Lipovetsky\altaffilmark{1,4} and Alexei Kniazev\altaffilmark{1}}
\affil{Special Astrophysical Observatory, Russian Academy of Sciences, Nizhny Arkhyz, Karachai-Circessia 357147, Russia; akn@sao.ru}

\author{J. Ward Moody}
\affil{Department of Physics \& Astronomy, Brigham Young University, Provo, UT 84602; jmoody@astro.byu.edu}

\author{Todd A. Boroson}
\affil{National Optical Astronomy Obs., P.O. Box 26732, Tucson, AZ 85726; tyb@noao.edu}

\author{Trinh X. Thuan}
\affil{Astronomy Department, University of Virginia, Charlottesville, VA 22903; txt@starburst.astro.virginia.edu}

\author{Yuri I. Izotov}
\affil{Main Astronomical Observatory, National Academy of Sciences of Ukraine, Goloseevo, Kiev 03680, Ukraine; izotov@mao.kiev.ua}

\author{Jos\'e L. Herrero}
\affil{BBN Technologies, Cambridge, MA 02140; jose@world.std.com}

\author{Lisa M. Frattare\altaffilmark{1}}
\affil{Space Telescope Science Institute, Baltimore, MD 21218; frattare@stsci.edu}

\altaffiltext{1}{Visiting Astronomer, Kitt Peak National Observatory. 
KPNO is operated by AURA, Inc.\ under contract to the National Science
Foundation.} 
\altaffiltext{2}{NSF Presidential Faculty Fellow.} 
\altaffiltext{3}{present address: Department of Physics \& Astronomy, Johns Hopkins University,
Baltimore, MD 21218; caryl@adcam.pha.jhu.edu.}
\altaffiltext{4}{Deceased 22 September 1996.} 

%\clearpage

\begin{abstract}
The KPNO International Spectroscopic Survey (KISS) is a new objective-prism
survey for extragalactic emission-line objects.  It combines many of the 
features of previous slitless spectroscopic surveys with the advantages of 
modern CCD detectors, and is the first purely digital objective-prism survey
for emission-line galaxies.  Here we present the first list of emission-line 
galaxy candidates selected from our red spectral data, which cover the spectral 
range 6400 to 7200 \AA.  In most cases, the detected emission line is 
H$\alpha$.  The current survey list covers a one-degree-wide strip located 
at $\delta$(1950) = 29$\arcdeg$~30$\arcmin$ and spanning the RA range 
12$^h$~15$^m$ to 17$^h$~0$^m$.  An area of 62.2 deg$^2$ is covered.  A total 
of 1128 candidate emission-line objects have been selected
for inclusion in the survey list (18.1 per deg$^2$).  We tabulate accurate
coordinates and photometry for each source, as well as estimates of the
redshift and emission-line flux and equivalent width based on measurements of 
the digital objective-prism spectra.  The properties of the KISS emission-line
galaxies are examined using the available observational data.
\end{abstract}

% The different journals have different requirements for keywords.  The
% keywords.apj file, found on aas.org in the pubs/aastex-misc directory, 
% contains a list of keywords used with the ApJ and Letters.  These are 
% usually assigned by the editor, but authors may include them in their 
% manuscripts if they wish. 

\keywords{galaxies: emission lines --- galaxies: Seyfert --- galaxies: starburst --- surveys}

%************************************************************************

\section{Introduction}

The KPNO International Spectroscopic Survey (KISS) is an ongoing objective-prism
survey which targets the detection of large numbers of extragalactic emission-line
sources.  The survey method is patterned after previous surveys for
these types of objects carried out with Schmidt telescopes and photographic
plates (e.g., Markarian 1967, Smith \etal 1976, MacAlpine \etal 1977, Pesch \& Sanduleak
1983, Wasilewski 1983, Markarian \etal 1983, Zamorano \etal 1994, Popescu \etal 
1996, Surace \& Comte 1998).
The primary characteristic that distinguishes KISS from these previous surveys
is that we utilize a CCD as our detector.  With the advent of large
format CCDs, the areal coverage possible with the combination of Schmidt
telescopes and CCDs makes digital surveys like KISS both possible and highly 
desirable.  The obvious advantages of CCDs over photographic plates include
much higher quantum efficiency, lower noise, good spectral response over the
entire optical portion of the spectrum, and large dynamic range.  In addition,
CCDs give us the ability to use automated selection methods to detect 
emission-line galaxies (ELGs), and allow us to quantify the selection function 
and completeness limit directly from the survey data.  The combination of 
increased depth and large areal coverage is a powerful one, and allows for substantial 
improvements compared to the previous photographic surveys listed above.

The primary goal of KISS is to produce a high-quality survey whose selection
function and completeness limits can be accurately quantified so that 
the resulting catalog of ELGs will be useful for a broad range of studies
requiring statistically complete galaxy samples.  We also want to reach
substantially deeper than previous objective-prism surveys.

A complete description of the survey method employed for KISS is given
in the first paper in this series (Salzer \etal 2000, hereafter Paper I).
As described there, the first survey strip was observed in two distinct
spectral regions.  One covered the blue portion of the spectrum
(4800 -- 5500 \AA), while the second covered the wavelength range 6400 -- 
7200 \AA\ in the red part of the spectrum.  The first blue survey list 
is given in Salzer \etal
(2001, hereafter KB1).  The present paper presents the initial KISS list
of H$\alpha$-selected ELG candidates.  In addition to listing the ELGs, we 
provide substantial observational data for each object.  This includes 
accurate photometry and astrometry for each source, as well as estimates of 
each galaxy's redshift, line flux, and equivalent width.  These data are 
used to examine the properties of the KISS ELGs in Section 4.

%************************************************************************

\section{Observations}

All survey data were acquired using the 0.61-meter Burrell Schmidt 
telescope\footnote{Observations made with the Burrell Schmidt of the
Warner and Swasey Observatory, Case Western Reserve University.
During the period of time covered by the observations described 
here, the Burrell Schmidt was operated jointly by CWRU and KPNO.}. 
The detector used for all data reported here was a 2048 $\times$ 
2048 pixel STIS CCD (S2KA).  This CCD has 21-$\micron$ pixels, which 
yielded an image scale of 2.03 arcsec/pixel.  The overall field-of-view 
was 69 $\times$ 69 arcmin, and each image covered 1.32 square degrees.   
The red survey spectral data were obtained with a 4$\arcdeg$ prism,
which provided a reciprocal dispersion of 24 \AA/pixel at H$\alpha$.
The spectral data were obtained through a special filter designed for
the survey, which covered the spectral range 6400 -- 7200 \AA\ (see
Figure 1 of Paper I for the filter transmission curve).

The survey data consist of the spectral images, obtained with the 
objective-prism on the telescope, plus direct images taken without
the prism through standard B and V filters.  The first survey strip 
was initially observed only in the blue spectral region (KB1).  The 
primary emission line 
detected in the blue survey data is [\ion{O}{3}]$\lambda$5007.  This first
blue survey consists of a contiguous strip of fields observed at a
constant declination ($\delta$(1950) = 29$\arcdeg$~30$\arcmin$).  
The R.A. range covered is 8$^h$~30$^m$ to 17$^h$~0$^m$.  This area was 
chosen to overlap completely the Century Redshift Survey (Geller \etal 
1997).  After experiments with spectral observations in the red
indicated great promise, we began reobserving the same fields, obtaining
spectral images with the prism and filter combination listed above.
The red survey used the existing direct images obtained for the blue
survey strip.  Limited observing time prevented us from reobserving
the entire strip in the red; only 54 of the 102 fields of the initial
blue strip were observed in the red.  These fields fall in the R.A.
range 12$^h$~15$^m$ to 17$^h$~0$^m$, but exclude three fields near the
middle of this strip from 14$^h$~30$^m$ to 14$^h$~45$^m$.  

The spectral data used for this paper were obtained during two observing
runs.  The first occurred in 1997 May 3--6, when objective-prism images of
20 survey fields were obtained,  and the second in 1997 June 5--12, when
an additional 34 fields were observed. The observations of the direct images 
will be described more completely in KB1; the same set of direct images was used
for both KB1 and the current red spectral survey.  The CCD used on the Burrell 
Schmidt was replaced after the 1997 June run, making additional red spectral 
observations of the existing survey fields less desirable.  Subsequent 
observations with the new CCD have focused on new survey regions.

All red spectral data consist of four 720 s exposures of each field, for
a total exposure time of 48 minutes per field.  The telescope is dithered 
by $\sim$10 arcsec between exposures in order to move sources off of
bad columns/pixels on the CCD.  Data processing procedures are detailed
in Paper I.  The analysis of the survey data is carried out using an
IRAF\footnote{IRAF is distributed by the National Optical Astronomy 
Observatories, which are operated by AURA, Inc.\ under cooperative 
agreement with the National Science Foundation.}-based software package 
written by members of the KISS team.  This package is described in 
Herrero \etal (2000).

%************************************************************************

\section{List 1 of the KPNO International Spectroscopic Survey}

\subsection{Selection Criteria}

The KISS reduction software selects ELG candidates by searching the
extracted objective-prism spectra for objects possessing 5$\sigma$
emission features.  That is, objects must possess one or more pixels 
with a flux level more than five times the total noise (Poissonian 
plus instrumental) above the neighboring continuum level in order to
be selected for inclusion.  This is the primary selection
criterion of the survey.  After the automated selection process is
completed, the candidate ELGs are checked manually by examining both
their extracted spectra and their appearance in the spectral and
direct images.  Many spurious sources are rejected at this stage.
These often consist of objects in which spikes have been introduced 
during spectral overlap corrections (see Herrero \etal 2000), bright 
stars which possess spectral breaks within the KISS bandpass that can 
mimic an emission line (e.g., M stars), and very faint objects with 
residual cosmic rays present.  Typically, $\sim$80 objects per field
are identified as ELG candidates by our automated software, but 
roughly 70\% are rejected as spurious during this checking process.
When the nature of a potential ELG is 
not clear from examination of its spectrum, we also consider other 
information to help assess whether or not it should be retained in 
the sample.  Useful parameters in this process include the B$-$V color 
and the object classification.  For example, objects classified as 
stars that also have very red colors are usually rejected as being 
spurious sources.  We stress that we do not require that an object 
be classified as a galaxy in order for it to be included in the 
final ELG list.  Many of the objects included in the current list 
are unresolved in the direct images (which have an effective 
resolution of $\sim$4--5 arcsec).

The visual examination of the spectral images also yields a number of 
objects that are missed by the automatic selection software but which 
do possess 5$\sigma$ emission features.  These are typically objects 
with lines very close to the red end of the spectra, or bright galaxies 
with lower equivalent width lines where the continuum fitting process 
overestimates the continuum level slightly and hence underestimates the 
line strength.  These objects are flagged manually as ELGs in the KISS 
tables.  For the current survey list, 17.8\% of the objects were added 
to the final catalog during this visual search.  The combination of the 
automatic software and this visual checking ensures that the sample is 
quite complete for objects with $\ge$5$\sigma$ lines.  A formal assessment
of the survey completeness will be presented in Gronwall \etal (2000).

The 5$\sigma$ threshold level for inclusion in the survey is a 
conservative value, one that was arrived at after substantial
testing.  While we wished to survey as deeply as
possible, and include as many potential ELGs as possible, we also
placed a high premium on the quality of the final survey lists.
Early tests involving follow-up spectroscopy carried out on fields 
where objects were selected to lower thresholds showed that 5$\sigma$
detections were nearly always real sources, while objects between 
4$\sigma$ and 5$\sigma$ tended to be real but also included a fair
number ($\sim$25\%) of spurious sources.  Below 4$\sigma$ the fraction 
of spurious sources increased dramatically.  Hence, we selected 5$\sigma$ 
as the detection threshold for the survey.  Objects with emission lines 
between 4$\sigma$ and 5$\sigma$ are also flagged in the database 
tables and retained as possible ELGs.  However, this sample of
$<$5$\sigma$ sources is not statistically complete, and hence is
not included in the main survey lists.  Rather, we will include 
these additional lower probability sources in a secondary list
of ELG candidates which should be thought of as a supplement to the
main KISS catalog.

\subsection{The Survey}

The list of ELG candidates selected in the first red survey is presented 
in Table 1.  Because of the nature of the survey data, we are able to 
include a great deal of useful information about each source, such as
accurate photometry and astrometry and estimates of the redshift
and emission-line flux and equivalent width.  Only the first page of the
table is printed here; the complete table is available in the electronic
version of the paper.

The contents of the survey table are as follows.  Column 1 gives a running 
number for each object in the survey with the designation KISSR XXXX, where 
KISSR stands for ``KISS red" survey.  This is to distinguish
it from the blue KISS survey (KB1).  Future red survey lists will 
continue with this numbering scheme.  Columns 2 and 3 give the object
identification from the KISS database tables, where the first number
indicates the survey field (FXXXX), and the second number is the ID number 
within the table for that galaxy.  This identifier is necessary for locating
the KISS ELGs within the survey database tables.  Columns 4 and 5 list
the right ascension and declination of each object (J2000).  The formal
uncertainties in the coordinates are 0.25 arcsec in RA and 0.20 arcsec in
declination.  Column 6 gives the B magnitude, while column 7 lists the 
B$-$V color.  For brighter objects the magnitude estimates have uncertainties
of typically 0.05 magnitude, increasing to $\sim$0.10 magnitude at B = 20.
Paper I includes a complete discussion of the precision of both the astrometry 
and photometry of the KISS objects.  An estimate of the redshift of each
galaxy, based on its objective-prism spectrum, in given in column 8.
This estimate assumes that the emission line seen in the objective-prism 
spectrum is H$\alpha$.  Follow-up spectra for $>$400 ELG candidates
shows that this assumption is correct in the vast majority of cases.
Only 8 ELGs with follow-up spectra are high redshift objects where a 
different line (typically [\ion{O}{3}] and/or H$\beta$) appears in the
objective-prism spectrum.  For objects with redshifts above z = 0.07 the 
observed value is corrected as described in Paper I.  The formal uncertainty
in these redshift estimates is $\sigma_z$ = 0.0028.  Columns 9 and 10
list the emission-line flux (in units of 10$^{-16}$ erg/s/cm$^2$) and
equivalent width (in \AA) measured from the objective-prism spectra.
The calibration of the fluxes is discussed in Section 4.1.2.  These
quantities should be taken as being representative estimates only.
A simple estimate of the reliability of each source, the quality flag (QFLAG),
is given in column 11.  This quantity, assigned during the line
measurement step of the data processing, is given the value of 1 for
high quality sources, 2 for lower quality but still reliable objects,
and 3 for somewhat less reliable sources.  Column 12 gives alternate identifications
for KISS ELGs which have been cataloged previously.  This is not an
exhaustive cross-referencing, but focuses on previous objective-prism
surveys which overlap part or all of the current survey area: Markarian (1967), 
Case (Pesch \& Sanduleak 1983), Wasilewski (1983), and UCM (Zamorano \etal 1994).  
Also included are objects in common with the {\it Uppsala General Catalogue of 
Galaxies} (UGC, Nilson 1973).

A total of 1128 ELG candidates are included in this first list of KISS
galaxies.  The total area covered by the survey is 62.2 deg$^2$, meaning
that there are 18.1 KISS ELGs per deg$^2$.  This compares to the surface 
density of 0.1 galaxies per deg$^2$ from the Markarian survey, and 0.56 
galaxies per deg$^2$ from the UCM survey; the present survey is much deeper!  
Of the total, 578 were assigned quality values of QFLAG = 1 (51.2\%), 388 have 
QFLAG = 2 (34.4\%), and 162 have QFLAG = 3 (14.4\%).  Based on our follow-up spectra
to date, 100\% (263 of 263) of the sources with QFLAG = 1 are {\it bona fide}
emission-line galaxies, compared to 93\% (111 of 119) with QFLAG = 2 and
only 61\% (28 of 46) with QFLAG = 3.  The properties of the KISS galaxy sample
are described in the next section.

\begin{figure*}[htp]
%\plotone{find2.eps}
\vskip-0.5in
\epsfxsize=6.8in
\hskip0.5in
\epsffile{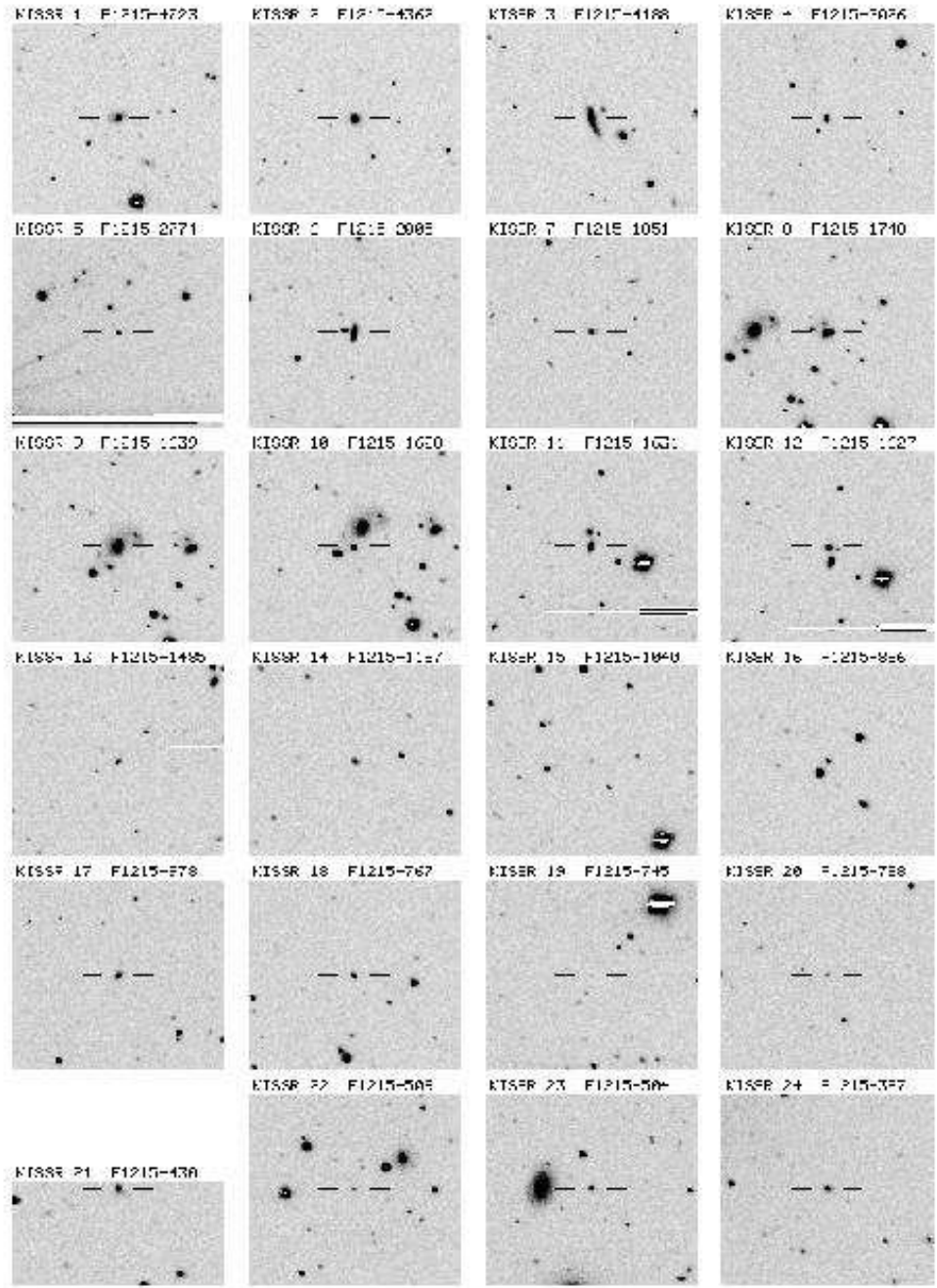}
\figcaption[fig1.eps]{Example of finder charts for the KISS ELG candidates.
Each image is 4.5 $\times$ 4.0 arcmin, with N up, E left.  These finders are 
generated from the direct images obtained as part of the survey.  In all 
cases the ELG candidate is located in the center of the image section 
displayed, and is indicated by the tick marks.\label{fig:find1}}
\end{figure*}

\begin{figure*}[htp]
\hskip0.5in
%\vskip-0.5in
\epsfxsize=7.0in
\epsffile{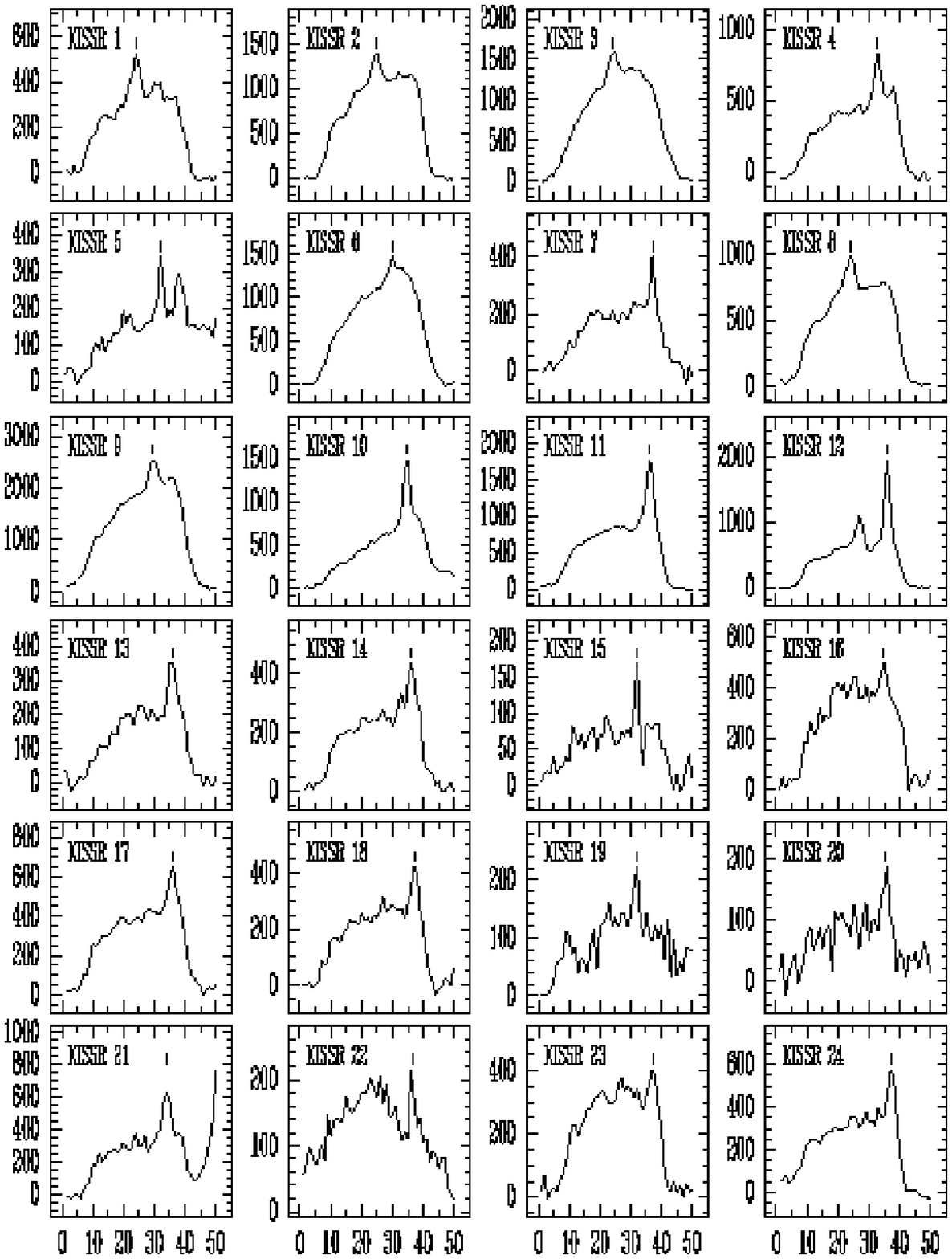}
%\epsfbox[75 85 540 705]{fig2.eps}
\figcaption[fig2.eps]{Plots of the objective-prism spectra for 24 KISS ELG
candidates.  The spectral information displayed represents the extracted
spectra present in the KISS database tables.  The location of the putative 
emission line is indicated.\label{fig:spec1}}
\end{figure*}

Figure~\ref{fig:find1} shows an example of the finder charts for the KISS 
ELGs.  These are generated from the direct images obtained as part of the
survey.  Figure~\ref{fig:spec1} displays the extracted spectra derived
from the objective-prism images for the first 24 ELGs in Table 1.  Finder
charts and spectral plots for all 1128 objects in the KISS survey are
available in the electronic version of this paper. 

A supplementary table containing an additional 189 ELG candidates is included 
in the appendix of this paper (Table 2).  These galaxies are considered to be lower
probability candidates, having emission lines with strengths between 4$\sigma$
and 5$\sigma$.  Follow-up spectra of a limited number of these objects
suggests that $\sim$75\% are real ELGs, compared to $\sim$94\% for the main
KISS catalog.  These additional galaxies do not constitute a statistically
complete sample, and should therefore be used with caution.  However, there
are likely many interesting objects contained in this supplementary list, 
hence we include them for the sake of completeness.

%************************************************************************

\section{Properties of the KISS ELGs}

A large amount of important observational information is available for
each KISS ELG candidate from the survey data.  This is due to the
survey method employed for KISS, which includes the acquisition of
photometrically-calibrated B and V direct images in addition to the 
objective-prism spectra.  Furthermore, the digital nature of the spectral 
data allows us to more readily measure quantitative information from the
spectra than was possible in previous photographic objective-prism
surveys.  Hence, a fairly complete picture of the survey constituents
can be developed without the need for follow-up observations.  Of course,
the spectral information available from the objective-prism data is
extremely coarse.  One cannot, for example, distinguish between active
galactic nuclei and star-forming galaxies based on the survey data alone.  
Follow-up spectra will be required in order to gain a more complete 
understanding of the nature of the KISS ELGs.  Still, the available data 
can be used to investigate the properties of the survey constituents.

\subsection{Observed Properties}

\subsubsection{Magnitude \& Color Distributions}

\begin{figure*}[ht]
\epsfxsize=6.0in
\hskip0.5in
\epsffile{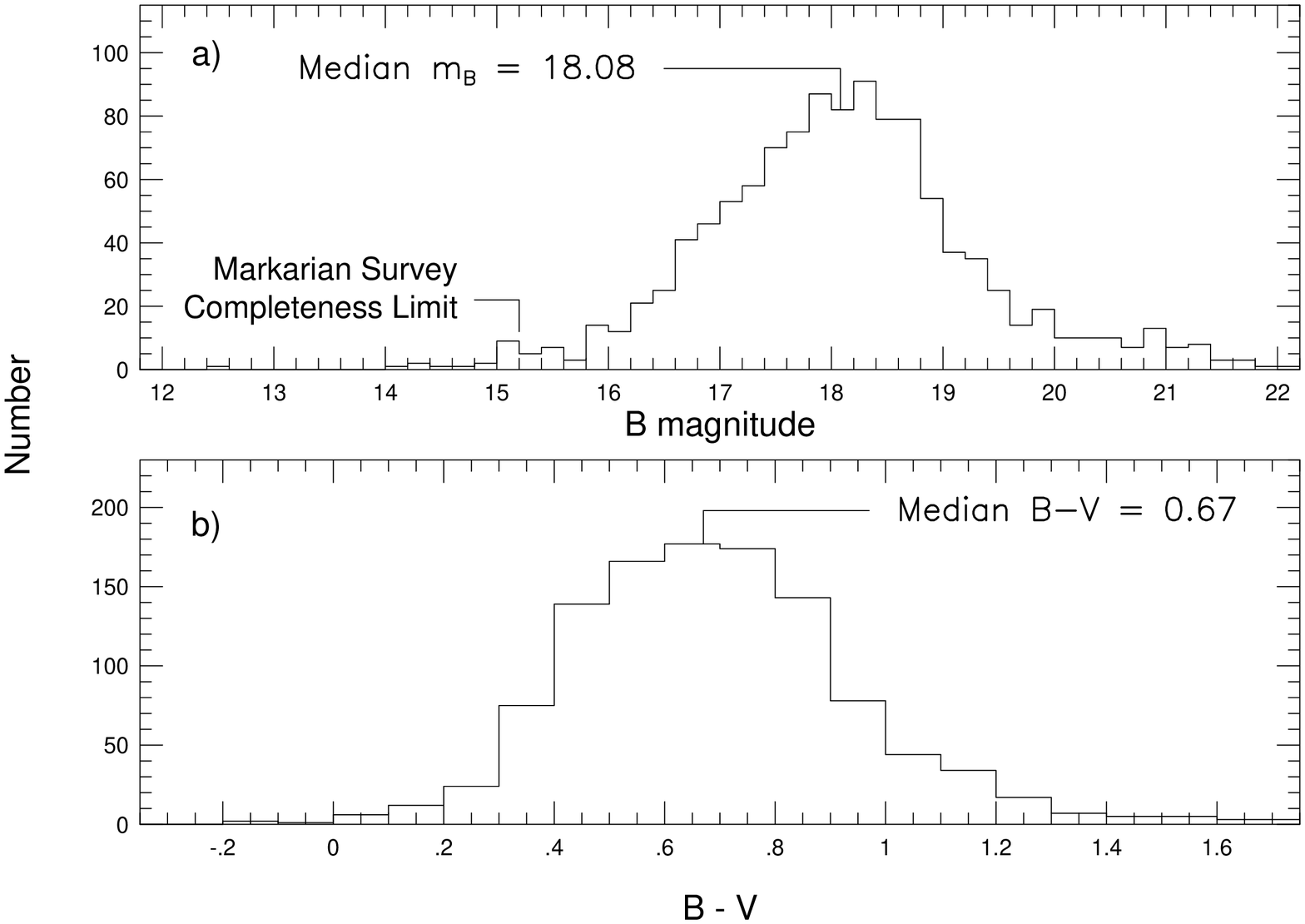}
\figcaption[fig3.eps]{(a) Distribution of B-band apparent magnitudes for the
1128 ELG candidates in the first H$\alpha$-selected KISS survey list.  The median 
brightness in the KISS sample is B = 18.08, with 7\% of the galaxies having 
B $>$ 20.  Also plotted, for comparison, is the completeness limit of the 
Markarian survey. (b) Histogram of the B$-$V colors for the 1128 ELG candidates.  
The median color of 0.67 is indicated. \label{fig:appmag}}
\end{figure*}

Figure~\ref{fig:appmag}a plots the histogram of B magnitudes for the 1128 KISS 
ELGs in the current survey list.  The median apparent magnitude is 18.08.  This
compares to median apparent magnitudes of B = 16.9 for the [\ion{O}{3}]-selected
Michigan (UM) survey (Salzer \etal 1989) and B $\approx$ 16.1 for the 
H$\alpha$-selected UCM survey (P\'erez-Gonz\'alez \etal 2000).  Also indicated 
in the figure is the completeness limit of the Markarian survey, B = 15.2 
(Mazzarella \& Balzano 1989).  Clearly, the survey goal of creating a deep
sample of ELGs has been achieved.

The distribution of B$-$V colors exhibited by the KISS sample is shown
in Figure~\ref{fig:appmag}b.  The median color of 0.67 is indicated.  This
median value, which is the typical color of an Sb galaxy (Roberts \& Haynes 1994),
is significantly redder than the median of B$-$V = 0.54 for 
the UM survey (Salzer \etal 1989).  This is most likely due to the differing
selection criteria for the two surveys.  The [\ion{O}{3}]-selected UM
survey preferentially detected lower luminosity ELGs with low amounts of
reddening.  While the H$\alpha$-selected KISS sample contains a large
number of dwarf galaxies as well (see Section 4.2.1), it has a larger
proportion of more luminous starburst galaxies and AGN.  There is also a
significantly weaker bias against heavily reddened ELGs with KISS compared
to the UM or other blue-selected surveys.  Hence the color range exhibited
by the KISS sample is closer to that of the overall galaxian population.
The color distribution appears to be similar to that of the UCM galaxies
(P\'erez-Gonz\'alez \etal 2000), which have a mean B$-$r color of 0.71.
This mean value is comparable to the mean color of an Sbc galaxy
(Fukugita \etal 1995).

It is difficult at this time to assess the significance of the extended tail 
of very red ELGs.  A total of 50 KISS galaxies in the current list possess 
B$-$V colors $>$ 1.2 (i.e., redder than any normal galaxy).  Most of these 
objects are very faint: the median apparent magnitude of these red objects 
is B = 20.6.  At these magnitudes the uncertainties in the photometry are 
substantial (typical errors in the B$-$V color are 0.2 to 0.3 magnitude).
Furthermore, the proportion of spurious sources increases at these faint magnitudes.
Follow-up spectra will be required to confirm the nature of these very 
red objects.  During our early spectroscopic follow-up observations, 14
of these red ELG candidates were observed, and only five have been confirmed
as ELGs.  The remainder are stars or faint galaxies with no emission lines.
Hence, many of these red candidates are probably false detections.
Of the five true ELGs mentioned above, three are Seyfert galaxies, which
suggests that this red population of KISS objects harbors some 
interesting objects.

\subsubsection{Line Strength Distributions}

\begin{figure*}[ht]
\epsfxsize=6.0in
\hskip0.5in
\epsffile{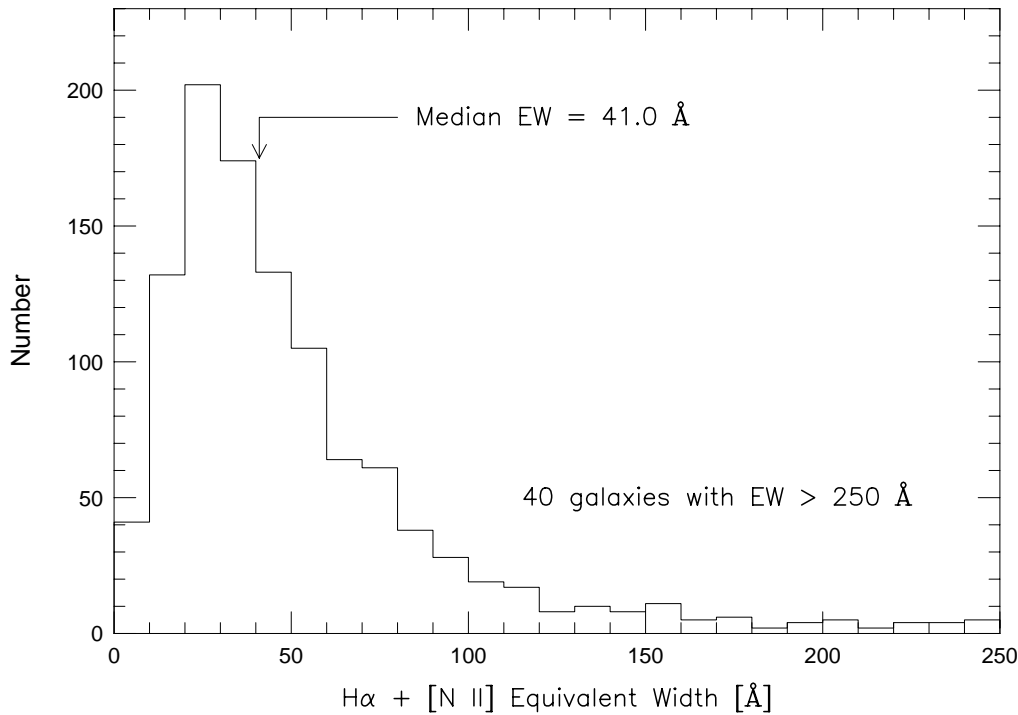}
\figcaption[fig4.eps]{Distribution of measured H$\alpha$ + [\ion{N}{2}]
equivalent widths for the KISS ELGs.  The median value of 41 \AA\ is 
indicated.  The measurement of equivalent widths from objective-prism 
spectra tends to yield underestimates of the true equivalent widths, so
these values should only be taken as estimates.  The survey appears to 
detect most sources with EW(H$\alpha$+[\ion{N}{2}]) $>$ 30 \AA.\label{fig:ew}}
\end{figure*}

As described in Paper I and Herrero \etal (2000), we are able to
measure the position and strength of the emission line that is seen
in the objective-prism spectrum of each KISS ELG.  The line strength
measurements, both line flux and equivalent width, yield useful 
quantitative information.  This allows us to use the survey
data to assess the completeness of the survey.  Since KISS is a
line-selected survey, its completeness limit must be defined in
terms of line strengths, not continuum apparent magnitudes
(Salzer 1989).  We discuss this issue further in Gronwall \etal (2000).

Given the low-dispersion nature of our spectra plus the
low spatial resolution of the images, the measured line strengths
are not as precise as those obtained from slit spectra.
In particular, the nature of the objective-prism spectra makes the
equivalent width estimate highly uncertain, due to the variable amount
of underlying continuum flux present at the location of the line.
Since light from all portions of the object gets dispersed by the prism,
the location occupied by the emission line in the spectrum can have
continuum contributions from a large area of the galaxy.  Hence, for
extended sources, the equivalent width measured from the survey spectra
will tend to be significantly underestimated.  This is not an important
issue for compact objects.  Partially compensating for this effect is the
fact that our extracted spectra, which sum the flux over four pixels 
($\sim$8$\arcsec$), tend to include all of the line emission
from each source.  This is in contrast to slit spectroscopy,
where the 1--2$\arcsec$ slits typically employed can miss a large
fraction of the line emission.  Thus, the KISS line fluxes should yield 
fairly accurate estimates of the total H$\alpha$ + [\ion{N}{2}] emission.

The distribution of equivalent widths seen in our survey galaxies 
is plotted in Figure~\ref{fig:ew}.  We assume that the line we are 
measuring is H$\alpha$ + [\ion{N}{2}].  Based on our follow-up spectra to 
date (see Section 3.2), this appears to be a reasonable assumption.
Due to our low spectral resolution, the [\ion{N}{2}]$\lambda\lambda$6584,6548
lines are always seen blended with H$\alpha$; there is no way to determine
the contribution from the individual lines using our survey data.  We note
in passing that the [\ion{S}{2}]$\lambda\lambda$6731,6717 doublet is
well resolved from the H$\alpha$ + [\ion{N}{2}] complex, and is often seen
in the objective prism spectra of strong-lined objects.  The median 
equivalent width found from the current sample of ELGs is
41.0 \AA.  As can be seen in Figure~\ref{fig:ew}, the vast majority of the
galaxies have equivalent widths less than 100 \AA.  The distribution of
EWs peaks in the 20--30 \AA\ bin, suggesting that KISS is fairly complete
for objects with EWs above $\sim$30 \AA.

\begin{figure*}[ht]
\epsfxsize=6.0in
\hskip0.5in
\epsffile{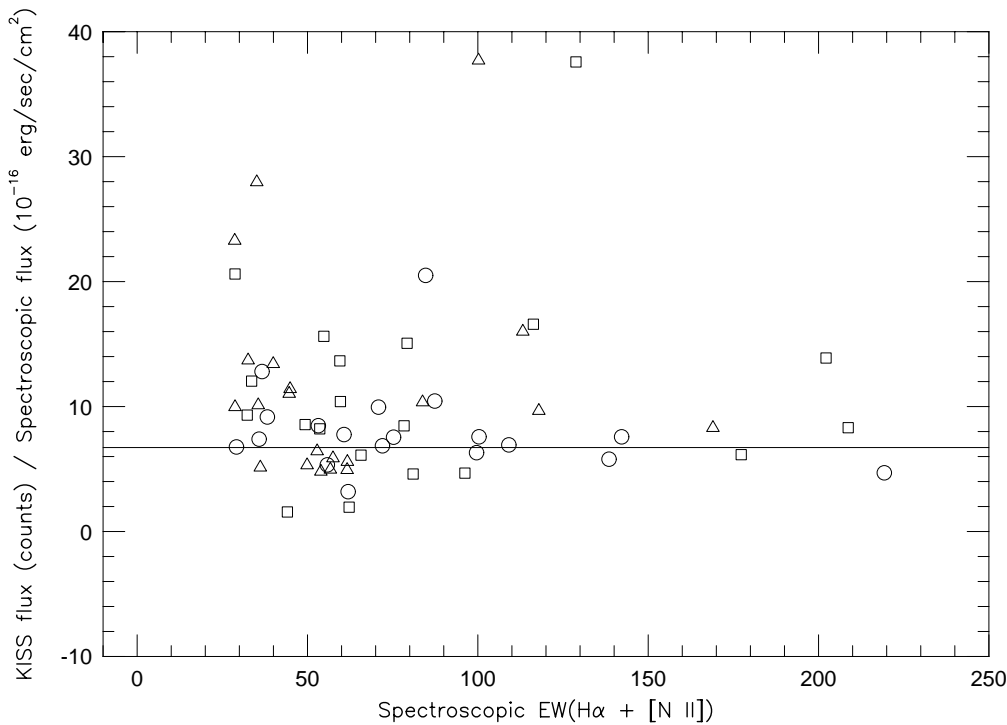}
\figcaption[fig5.eps]{Plot of the ratio of objective-prism flux (in counts) 
to spectroscopic flux {\it versus} H$\alpha$~+~[\ion{N}{2}] equivalent width measured from 
the follow-up spectra.  Data from three different observing runs on two 
different telescopes are plotted: square = MDM 2.4-m in May, 1998, triangle =
MDM 2.4-m in April, 1999, and circle = KPNO 2.1-m, May 1999.  See Gronwall
\etal (2001) for details of the spectroscopic observing runs.  The solid
line indicates the median ratio.  Five galaxies with EW $>$ 250 \AA\ lie off
the diagram to the right.\label{fig:lfluxcal}}
\end{figure*}

Calibrating the fluxes measured from the objective-prism spectra is crucial 
for scientific analyses such as determining the star formation rate
of the local universe (Gronwall \etal 2000), as well as for establishing 
the completeness limit of the survey.  The KISS line fluxes are calibrated 
in a two-step method.  First, the objective-prism spectra for each field are 
corrected for throughput variations and atmospheric extinction (see Herrero 
\etal 2000).  This places all line fluxes on the same {\it relative} flux 
scale.  Then, the fluxes are calibrated using information obtained from our 
early follow-up spectra.  Since the measured objective-prism line fluxes are
often of low precision, due to the coarse nature of the spectra, the goal
of the following calibration procedure was not necessarily to achieve
high precision.  Rather, an uncertainty of between 10 to 20\% was considered
acceptable.  

\begin{figure*}[ht]
\epsfxsize=6.0in
\hskip0.5in
\epsffile{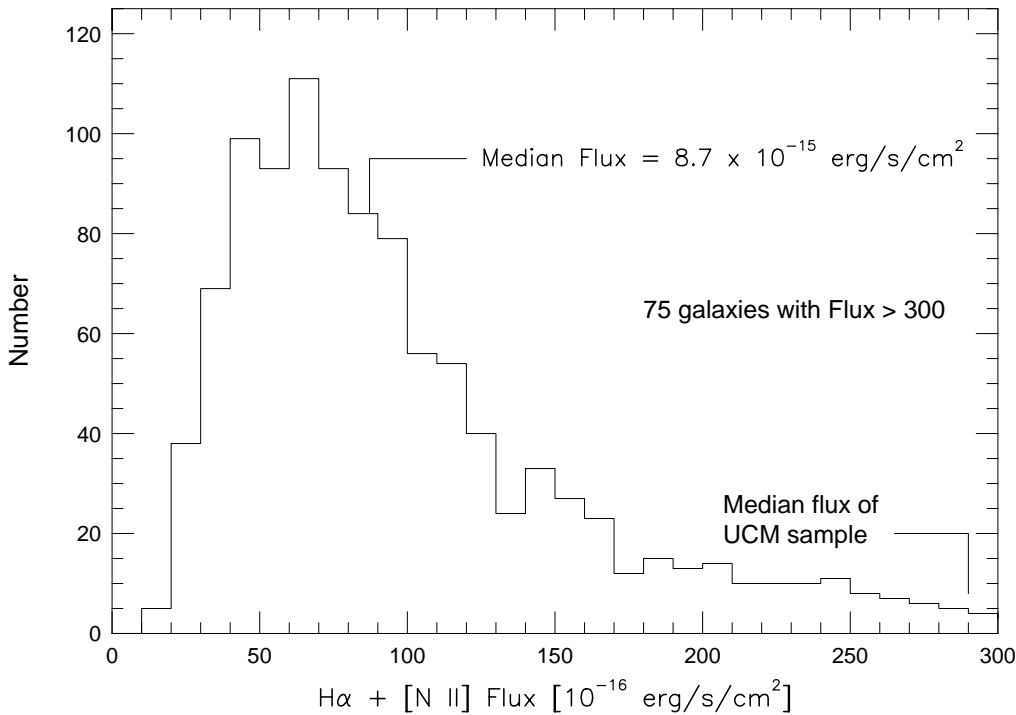}
\figcaption[fig6.eps]{Distribution of H$\alpha$ + [\ion{N}{2}] line fluxes
for the 1128 KISS ELGs included in the current survey list.  The median flux
level of both the KISS and UCM samples is indicated.\label{fig:lflux}}
\end{figure*}

We have obtained high-resolution follow-up spectra for 428 KISS candidates 
from the sample of 1128 presented in this paper.  The spectroscopic observations
will be described in Gronwall \etal (2001).  For the purposes of calibrating
the objective-prism spectra, we selected galaxies that are starbursting
(i.e., not AGN), that had been observed with a longslit spectrograph under 
photometric conditions, and for which high quality spectra were obtained.  
Galaxies observed through optical fibers were not used, since such spectra are
notoriously difficult to use for accurate spectrophotometry.  This limited our 
calibration sample to 65 galaxies.  All data used were obtained using slit
widths of either 1.7 or 2.0 arcsec.  This guarantees that we recorded most of
the emission-line flux from the sources, as long as the emission regions were
not spatially extended.  Since the fluxes measured from the 
objective-prism spectra are a combination of the  H$\alpha$ and [\ion{N}{2}]
lines, we used the fluxes from our slit spectra for the sum of these three
lines.  Figure~\ref{fig:lfluxcal} shows a plot of the ratio of objective-prism
flux (in counts) to spectroscopic flux {\it versus} the equivalent width 
measured from the follow-up spectra.  Data from three different observing runs 
on two different telescopes are plotted, and there are no obvious systematic 
trends in this ratio with either equivalent width or observing run.  There are, 
however, a number of galaxies with large ratios, particularly at low equivalent 
width.  A visual inspection of the survey images showed that these galaxies 
are of large angular extent and exhibit extended line emission.  The high flux 
ratios for these objects indicate that our longslit measurements do not include 
all of the H$\alpha$ emission from these sources.  Thus, we restricted our 
analysis to those galaxies with an objective-prism-to-spectroscopic flux ratio 
of less than 12.5 and an equivalent width greater than 40 \AA, leaving us with 
a calibration sample of 42 galaxies.  These galaxies all possess emission regions
that are essentially point sources.  The median ratio of this sample was 6.72; 
the mean was 6.91 with a standard deviation of 2.48 and an error in the mean of 
0.38.  We have adopted as our calibration value the reciprocal of the median value, 
or 0.1488 $\times 10^{-16}$ ergs/sec/cm$^{-2}$ per count.

To check our calibration, we performed a similar analysis using the 17 galaxies 
from the UCM survey with spectral data in Gallego \etal (1996) that fall 
within our survey region.  The KISS and UCM line fluxes are plotted in Figure 8 
of Paper I.  The resulting flux-ratio analysis for the UCM galaxies yields a median 
ratio of 6.62 and a mean ratio of 7.05, with a standard deviation of 1.09 and an 
error in the mean of 0.30.  Thus the value agrees quite well (within 1.5\%) with our 
derived calibration factor.   Given the necessarily coarse nature of the objective-prism
line fluxes, we feel that the accuracy of our flux calibration method (formal
error of 6\%) is quite acceptable.   

Figure~\ref{fig:lflux} displays the histogram of observed H$\alpha$+[\ion{N}{2}]
line flux values for the 1128 KISS ELGs.  The median value of 8.7 $\times$ 
10$^{-15}$ erg/s/cm$^2$ compares to that from the UCM sample of 2.9 $\times$ 
10$^{-14}$ erg/s/cm$^2$ (based on follow-up spectra of Gallego \etal 1996).
A galaxy with this latter flux level would fall at the extreme right edge of the 
figure, which illustrates the increased depth of KISS relative to the UCM survey
in terms of line flux.

\subsubsection{Redshift Distributions}

\begin{figure*}[ht]
\epsfxsize=6.0in
\hskip0.5in
\epsffile{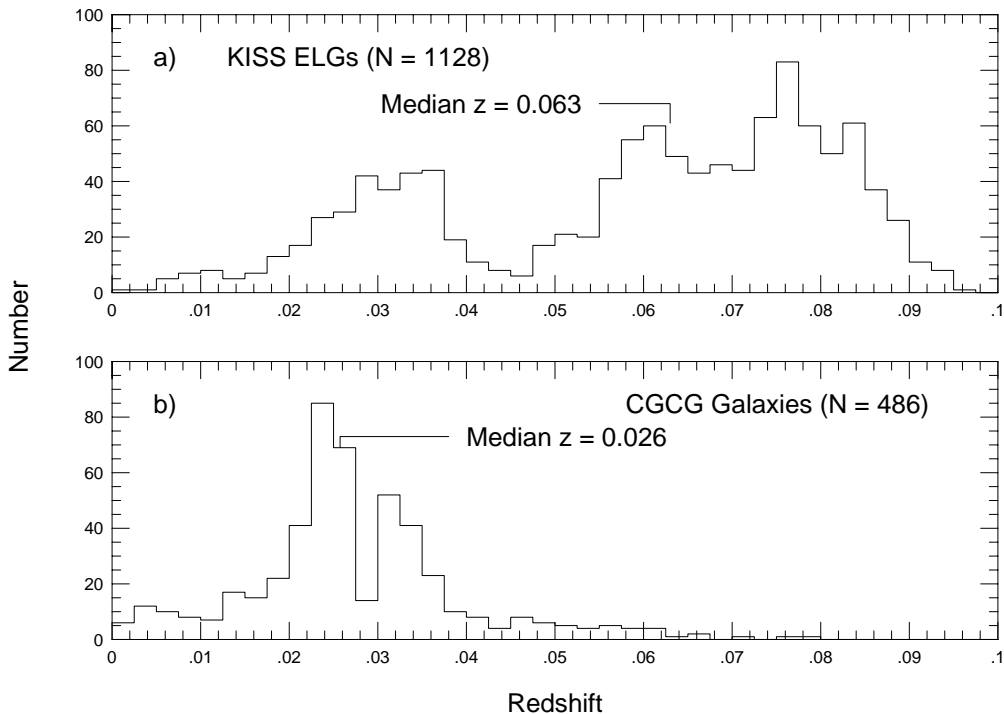}
\figcaption[fig7.eps]{Histograms showing the distribution of redshift for 
(a) the 1128 H$\alpha$-selected KISS ELGs and (b) the 486 ``normal" galaxies 
from the CGCG which are located in the same area of the 
sky.  The median redshift is indicated in both plots.  Note that the number 
of KISS ELGs continues to rise up to the cut-off of the filter used for the 
survey, indicating that the survey is volume-limited for the more luminous 
galaxies.  The deficit of ELGs between z = 0.038 and 0.054 is due to a large 
void. \label{fig:zhist}}
\end{figure*}

The other parameter that we measure from the objective-prism spectra is the 
redshift of each object.  The precision of the KISS redshifts is detailed in
Paper I, where we show that the formal uncertainty in the KISS redshifts is
0.0028 (830 \kms).  The distribution of the measured redshifts is illustrated in 
Figure~\ref{fig:zhist}.  The upper panel shows the KISS ELGs, while
the lower panel shows the distribution for 486 galaxies from Zwicky \etal
(1961--1968; hereafter CGCG).  The redshifts for the CGCG galaxies are taken
from Falco \etal (1999); this is essentially the portion of the CfA2 redshift 
survey that covers the same area on the sky as the KISS ELGs.  Because the 
surface density of the CGCG catalog is fairly low, we have actually included 
objects lying in a four-degree-wide declination strip, rather than the one 
degree covered by KISS.  The Falco \etal redshift catalog sample is complete 
to m$_B$ = 15.5.  This same comparison sample is used in the following section 
as well.

Compared to the CGCG sample, the KISS ELGs are detected in
large numbers to much higher redshifts.  The median redshift for the ELGs
is more than double that of the magnitude-limited CGCG galaxies.  There is
a large void present between z = 0.038 and 0.054 (see also Figure~\ref{fig:cone}).
Beyond this void, the number of KISS ELGs increases, while the CGCG galaxies
remain sparse.  Only a handful of luminous galaxies are present in the
CGCG sample beyond z = 0.04.  In contrast, the KISS ELGs increase in number
out to the redshift limit imposed by the survey filter.  The implications
of this are (1) that the survey technique employed by KISS would be sensitive to
galaxies at distances well beyond the distance limit imposed by the filter, if
the filter was either not used or was replaced with one that extended to
redder wavelengths, and (2) for the high-luminosity portion of the ELG luminosity 
function, the KISS sample is effectively volume-limited rather than flux-limited.

\subsection{Derived Properties}

\subsubsection{Luminosity Distribution}

\begin{figure*}[ht]
\epsfxsize=6.0in
\hskip0.5in
\epsffile{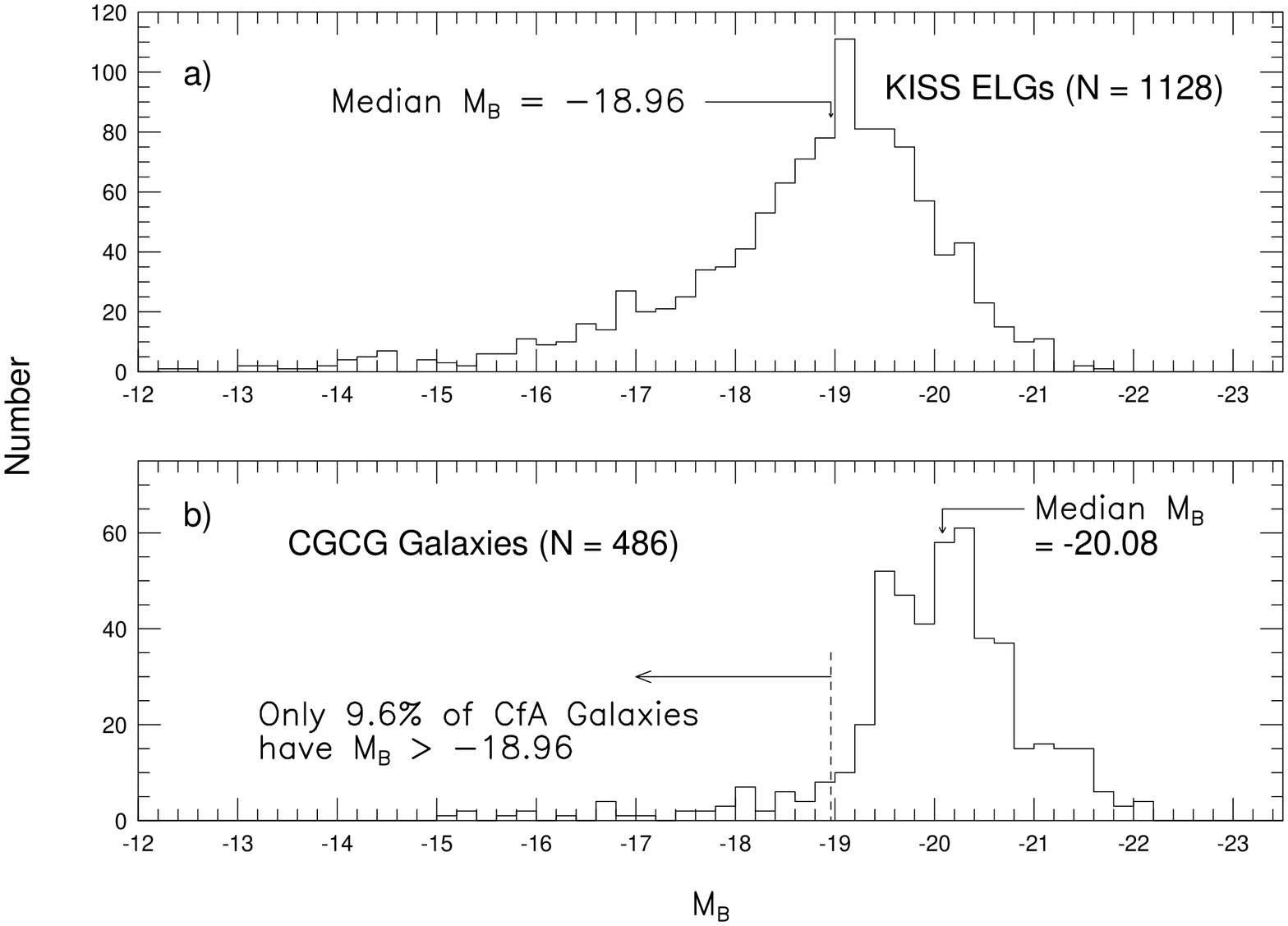}
\figcaption[fig8.eps]{Histograms showing the distribution of blue absolute 
magnitude for (a) the 1128 H$\alpha$-selected KISS ELGs and (b) the 486 
``normal" galaxies from the CGCG that are located in the 
same area of the sky.  The median luminosity of each sample is indicated.  The 
KISS ELG sample is made up of predominantly intermediate- and lower-luminosity 
galaxies, making this line-selected sample particularly powerful for studying 
dwarf galaxies. \label{fig:absmag}}
\end{figure*}

We compare the luminosities of the KISS ELGs with those of the CGCG galaxies
located in the same area of the sky in Figure~\ref{fig:absmag}.  Absolute
magnitudes are computed using the redshifts and apparent magnitudes listed
in Table 1 and assuming a value for the Hubble Constant of H$_o$ = 75 km/s/Mpc.
Corrections for Galactic absorption (A$_B$) have been applied by averaging the values
for all UGC galaxies in each survey field from the compilation of Burstein \&
Heiles (1984).  Since the majority of the survey strip is at high Galactic
latitude, this correction is typically small: 31 of the 54 fields (57\%) have 
A$_B$ $<$ 0.05, and 48 of 54 (89\%) have A$_B$ $<$ 0.10.  The maximum correction
of A$_B$ = 0.22 occurs in the easternmost survey field (F1655).
The median blue absolute magnitude of the KISS ELGs is $-$18.96, which is roughly
one magnitude fainter than M$^*$, the ``characteristic luminosity" parameter
of the Schechter (1976) luminosity function.  As seen in the lower portion of the
figure, the majority of the CGCG galaxies are more luminous than the KISS ELGs.  
The median absolute magnitude of the CGCG sample is $-$20.08 (i.e., very close 
to M$^*$).  Only 9.6\% of the CGCG galaxies have luminosities below the 
median KISS luminosity.

Figure~\ref{fig:absmag} reveals much about the KISS sample.  It is dominated
by intermediate- and low-luminosity galaxies.  The typical luminosity is
comparable to the Large Magellanic Cloud.  There is also a significant population 
of higher luminosity galaxies, but the high-luminosity end of the distribution is
truncated relative to the CGCG sample.  This is due to the volume-limited
nature of the luminous end of the sample.  Due to the filter-induced
cut-off in the redshift distribution, the KISS sample lacks the high
luminosity tail that is commonly seen in magnitude-limited samples.

Even though there are large numbers of lower luminosity ELGs present in
KISS, the proportion of dwarf star-forming systems is significantly lower
than in [\ion{O}{3}]-selected surveys like the UM survey.  The median
absolute magnitude for the UM ELGs is M$_B$ = $-$18.1 (Salzer \etal 1989).  
Due to two factors -- the way that the [\ion{O}{3}] line strength varies with 
metallicity, and the luminosity-metallicity relation -- [\ion{O}{3}]-selected 
surveys have a selection function that peaks near M$_B$ = $-$17.  More
luminous galaxies with starbursts tend to have weaker [\ion{O}{3}] lines.
In addition, [\ion{O}{3}]-selected surveys tend to be biased against 
luminous galaxies due to the higher amounts of reddening present in these
more metal-rich systems.  Since KISS selects by H$\alpha$, it does not
suffer from these biases.  Hence, the KISS sample should be a more
representative catalog of AGN and star-forming galaxies.  

Despite the difference in the median luminosities of the UM and KISS samples, 
one should not conclude that the KISS sample is deficient in dwarf galaxies 
relative to the UM survey.  Rather, the UM survey is missing a large fraction 
of the more luminous star-forming galaxies, which KISS recovers.  This can
be understood by consideration of the ELGs discovered in the first blue survey 
list (KB1).  The luminosity distribution of the KB1 sample is nearly identical 
to that of the UM ELGs of Salzer \etal (1989).  However, as discussed in Paper I,
there is a tremendous overlap between the KB1 and current samples.  In the regions
where they overlap, 92\% of the KB1 ELGs are also cataloged in the current paper
(87\% in the main catalog, and 5\% in the supplementary list).  Hence, the
H$\alpha$ survey technique is just as efficient at finding dwarf ELGs as is the
[\ion{O}{3}]-selection method.  In addition, the H$\alpha$-selected samples
detect many more higher-luminosity galaxies.

\subsubsection{Spatial Distribution}

\begin{figure*}[htp]
\epsfxsize=6.5in
\hskip0.5in
%\vskip0.1in
\epsffile{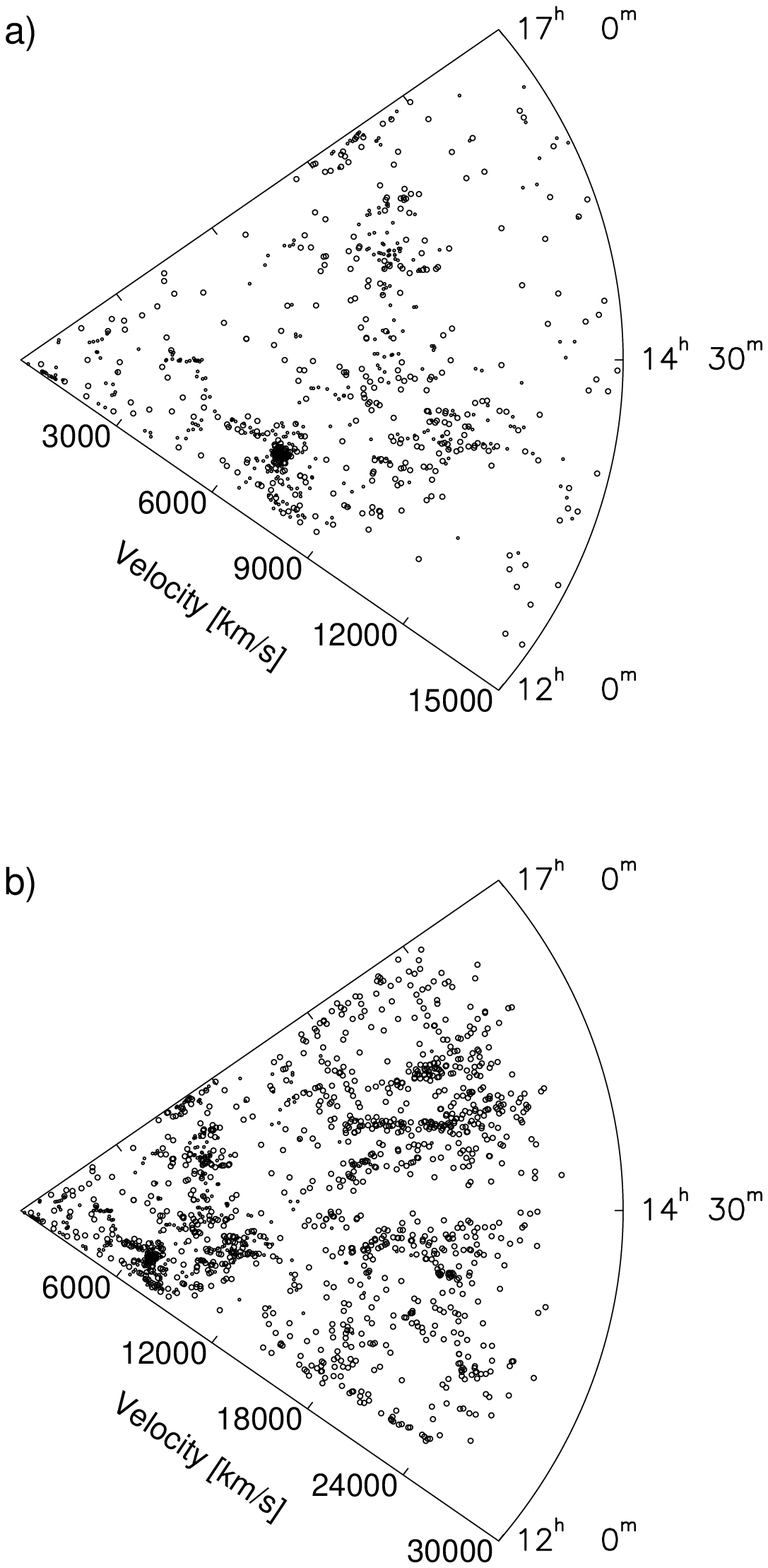}
%\null\vskip-8.0in
%\def\epsfsize#1#2{0.85#1}
%\epsfbox[125 70 535 415]{cone1.eps}
%\vskip-8.0in
%\def\epsfsize#1#2{0.8#1}
%\epsfbox[125 70 535 415]{cone1.eps} 
\figcaption[fig9.ps]{The spatial distribution of the KISS ELGs.  The CGCG
comparison sample is displayed as small dots, while the ELGs are larger, open
symbols.  (a) Velocities plotted out to 15,000 \kms.  The ELGs are seen to 
trace the large-scale structures defined by the CGCG galaxies at low redshift.
However, they exhibit the appearance
of being less tightly clustered.  A large number of ELGs are seen to fall
in voids.  (b) Velocities plotted out to 30,000 \kms.  At larger distances 
the ELGs appear to reveal several structures not visible in the shallower 
CfA2 redshift catalog.  The numbers of ELGs remains high out to 
$\sim$27,000 \kms, the point where the filter used for the survey truncates 
the sample.  The lack of objects in the ELG distribution near 14$^h$ 40$^m$ 
is caused by a 15$^m$ gap in the survey (see Section 2).
\label{fig:cone}}
\end{figure*}

Because of the depth and good redshift coverage of the KISS ELGs, it is
also relevant to consider the spatial distribution of the sample.  This is
illustrated in Figure~\ref{fig:cone}, which shows two cone diagrams for the
KISS and CGCG galaxies.  The redshifts plotted for the KISS galaxies are 
those from Table 1.  Figure~\ref{fig:cone}a shows all galaxies from both
samples out to 15,000 \kms (z $\le$ 0.05).
The general impression obtained from examination of the figure is that the
KISS ELGs tend to fall along the same large-scale structures seen at lower
redshifts in the CGCG galaxies.  However, the ELGs appear to be less tightly
confined to the structures than are the ``normal" galaxies.  Because of the 
limited precision of the KISS redshifts ($\pm$830 \kms; see Paper I), it is 
possible that the appearance of lower clustering is an artifact of the data.
However, the large numbers of galaxies located well into some of the voids
present within this volume of space suggest that the lower level of clustering is
in fact real.  Previous studies of the spatial distribution of ELG samples
have found similar results (Salzer 1989, Rosenberg \etal 1994, Pustil'nik \etal
1995, Popescu \etal 1997, Lee \etal 2000).  The lower level of clustering seen
in Figure~\ref{fig:cone}a is due primarily to lower-luminosity ELGs.  We are 
currently analyzing the relative clustering strengths of the ELGs and CGCG 
galaxies, using more accurate redshifts from our follow-up spectra when 
available (Lee, Salzer \& Gronwall 2001).

Figure~\ref{fig:cone}b shows the galaxian distribution out to z = 0.10.
At redshifts beyond about 12,000 \kms, the CGCG sample thins out drastically,
and can no longer be used to delineate large-scale structure.
The KISS galaxies, however, are found in great numbers out to $\sim$27,000 \kms,
beyond which the survey is truncated abruptly by the filter.  The ELGs
in the cone diagram appear to indicate the presence of a number of filaments
and voids at these higher redshifts.  In fact, a comparison of 
Figure~\ref{fig:cone}b with the similar diagram for the Century Redshift
Survey (Geller \etal 1997) reveals that these apparent structures are 
real.  The ELGs map out large-scale structures, albeit only roughly, 
out to z = 0.09 without the need for follow-up spectra!  While the limited 
velocity resolution, together with the tendency for the ELGs to be less clustered,
prevents the use of these data for detailed definition of 
large-scale structures at these redshifts, they certainly provide a reasonable 
picture of the gross details.  Furthermore, since the KISS galaxies are selected 
due to their strong line emission, follow-up spectroscopy of these faint galaxies 
is relatively easy.  Thus, using ELGs to trace out the main features in the spatial 
distribution of galaxies is an efficient use of telescope time. 

\subsection{Comparison with Previous Surveys}

Table 1 lists cross-references for KISS ELGs which are also cataloged in
other surveys for active and star-forming galaxies.  Four major surveys
overlap the first red KISS strip: Markarian (1967), Case (Pesch \& Sanduleak
1983), Wasilewski (1983), and UCM (Zamorano \etal 1994).  It is instructive
to consider the degree of overlap between these previous photographic
surveys and KISS.

Two of the four surveys mentioned above are strictly line selected.  The 
Wasilewski survey is [\ion{O}{3}] selected, while UCM is H$\alpha$ selected.
The Markarian survey is UV-excess selected, while the Case survey is a hybrid
of the two types: both line and UV-excess selected.  One might then expect
that KISS has the highest degree of overlap with the Wasilewski and UCM
surveys.  This is in fact the case.  There are 8 Wasilewski galaxies within
the area surveyed by KISS, and all 8 are recovered by KISS.  Twenty five 
UCM galaxies reside in the KISS area, but follow-up spectroscopy by Gallego found
that seven of these do not possess emission lines.  Hence, there are only 18 
UCM galaxies with emission in the KISS area, and all 18 are found by KISS.  
Oddly enough, one of the seven UCM non-ELGs was recovered by KISS 
(KISSR 185 = UCM 1300+2959); a follow-up spectrum for this object has not
yet been obtained.

While our first H$\alpha$-selected KISS list includes 100\% of the ELGs from the previous 
line-selected surveys, it did not do as well with the UV-excess-selected surveys.  
There are seven Markarian galaxies in the survey area.  Only four of these were
recovered by KISS in the main survey (i.e., Table 1), while two additional objects 
are included in the supplementary list of 4$\sigma$ to 5$\sigma$ objects.  The one 
Markarian survey galaxy not detected at all by KISS is Mrk 655 (B = 15.5).  Examination 
of its objective-prism spectrum shows no hint of an emission line.  The redshift of
Mrk 655 is less than the limit set by our survey filter, meaning that we did not miss
it because the H$\alpha$ line is redshifted beyond our spectral window.  Since the 
Markarian survey is not line-selected, it is perhaps no surprise that not all
of the objects from this survey were found by KISS.   Of the 55 Case galaxies 
in the KISS area, 39 were recovered by KISS (72\%).  We examined the 16 Case
galaxies not recovered by KISS, and found that 11 are listed in the Case Survey 
papers as being color-selected, while the remaining five are all listed as having 
questionable (w? or w:) line detections.  Spectroscopy of 61 Case galaxies by 
Salzer \etal (1995) in the red portion of the spectrum found that 31\% had 
H$\alpha$ equivalent widths less than 30 \AA.  Hence, it would appear that 
the non-detected Case galaxies are simply the weak-lined subsample of the 
Case survey, to which KISS is not sensitive.

%************************************************************************

\section{Summary}

We present the first list of emission-line galaxy candidates from the KPNO 
International Spectroscopic Survey.  KISS is an objective-prism survey, similar 
in nature to a number of important surveys carried out in the past using Schmidt
telescopes and photographic plates.  The crucial improvement incorporated into
KISS is the use of a CCD as the detector.  This allows us to detect fainter ELGs 
than previous surveys.  Our specially chosen filters also allow us to detect
emission lines out to higher redshifts than did most previous line-selected
surveys.  The combination of higher sensitivity, lower noise, and larger
volumes surveyed yield huge improvements in the depth of the resulting survey.  
KISS finds 181 times more AGN and starburst galaxy candidates per unit area than 
did the Markarian (1967) survey, and 32 times more than the UCM survey (Zamorano 
\etal 1994).

This first installment of KISS includes 1128 ELG candidates selected from
54 red survey fields covering a total of 62.2 deg$^2$.  The KISS catalog has 
a surface density of over 18 galaxies per deg$^2$.  The primary emission
line we are sensitive to is H$\alpha$.  The survey follows a narrow strip 
across the sky at a declination of $\delta$(1950) = 29$\arcdeg$~30$\arcmin$ 
and spanning the RA range 12$^h$~15$^m$ to 17$^h$~0$^m$.  This region was 
chosen to overlap the Century Redshift survey (Geller \etal 1997).  For each 
object in the catalog we tabulate accurate equatorial coordinates, B \& V 
photometry, and estimates of the redshift and line strength measured from the 
objective-prism spectra. Also provided are finder charts and extracted spectral 
plots for all galaxies.  In addition to the main survey list, we include a 
supplementary list of 189 ELG candidates with weaker (lower significance) 
emission lines.  

Since the survey data themselves provide such a large amount of observational
data for each KISS ELG, we are able to develop a fairly complete picture of 
the survey constituents even before follow-up spectral information
is available.  The median apparent magnitude of the sample is B~=~18.08, which
is substantially fainter than previous ELG surveys.  Objects fainter than
B~=~20 are routinely cataloged.  Measurement of the line strengths in the 
objective-prism spectra show that KISS is sensitive to objects with H$\alpha$ + [\ion{N}{2}]
equivalent widths of less than 20 \AA, and that most objects with EW $>$ 30 \AA\ 
are detected.  The median line flux of the KISS sample is more than
three times lower than that of the UCM survey (Gallego \etal 1996).  The
luminosity distribution of the KISS ELGs is dominated by intermediate- and
low-luminosity galaxies, although luminous AGN and starbursting galaxies are
also represented.  The median absolute magnitude of M$_B$ = $-$18.96 is
characteristic of a small spiral galaxy or large Magellanic irregular, and
underscores the fact that strong-lined galaxies of the type cataloged by KISS
tend to be less luminous than the types of objects found in more traditional
magnitude-limited samples.

Despite the fact that one can learn a great deal about each KISS object from
the survey data alone, detailed follow-up spectra are still required to
get a more complete picture.  For example, one cannot distinguish between
AGN and star-formation activity in the KISS galaxies based on the objective-prism
spectra.  Further, the redshifts derived from the KISS spectral data are too
coarse to be used in detailed spatial distribution studies (e.g., Lee, Salzer
\& Gronwall 2001).  We are in the process of obtaining spectra for a large
number of KISS ELGs in order to better assess the nature of the individual
galaxies, as well as to allow for a wide range of science applications, many
of which are outlined in Paper I.

\acknowledgments

We gratefully acknowledge financial support for the KISS project through
NSF Presidential Faculty Award to JJS (NSF-AST-9553020), which was instrumental
in allowing for the international collaboration.  Additional support for
this project came from NSF grant AST-9616863 to TXT, and from Kitt Peak National
Observatory, which purchased the special filters used by KISS.  Summer research 
students Michael Santos, Laura Brenneman, and Erin Condy, supported by the Keck 
Northeast Astronomy Consortium student exchange program, helped to reduce the 
survey data presented in the current paper.  We are grateful to Vicki Sarajedini 
and Laura Chomiuk, who assisted in so many ways during the final production of 
this paper, and to Katherine Rhode for her critical reading of the manuscript.  
Several useful suggestions by the anonymous referee helped to improve the
presentation of this paper.  We thank the numerous colleagues with whom we have 
discussed the KISS project over the past several years, including Jes\'us Gallego, 
Rafael Guzm\'an, Rob Kennicutt, David Koo, and Daniel Kunth.  Finally, we wish to 
thank the support staff of Kitt Peak National Observatory for maintaining the 
telescope and instrument during the early years of the project, and the Astronomy
Department of Case Western Reserve University for taking over this role after 1997.

%************************************************************************

%\clearpage
\appendix
\section{Supplementary Table of 4$\sigma$ Objects}

As explained in Section 3, the main selection criterion used to decide whether
or not an object is included in the KISS catalog is the presence of a 5$\sigma$
emission feature in its spectrum.  Because of the high sensitivity of the survey
data, many objects were detected with emission lines which were slightly
weaker than this level.  We made the decision to exclude such objects from the
main survey lists, in order to preserve the statistically complete nature of the
sample.  It was felt that the high degree of reliability of the sample would
be compromised somewhat if these objects were included.  However, rather than 
ignore these weaker-lined ELG candidates entirely, we are publishing them in
a supplementary table.

Listed in Table 2 are 189 ELG candidates that have emission lines detected
at between the 4$\sigma$ and 5$\sigma$ level.  The format of Table 2 is the
same as for Table 1, except that the objects are now labeled with KISSRx
numbers (`x' for extra).  The full version of the table, as well as finder 
charts for all 189 KISSRx galaxies, are available in the electronic version 
of the paper. 

\placetable{table:tab2}

The characteristics of the supplementary ELG sample are similar to those of the
main survey ELGs, although with some predictable differences.  The median
H$\alpha$ equivalent width is 19.3 \AA, a factor of two below the value for the
main sample.  The KISSRx galaxies are somewhat fainter (median B magnitude
of 18.6) and significantly redder (median B$-$V = 0.80).  Their median redshift
is slightly higher than that of the main sample (0.070), and their median
luminosity is slightly lower ($-$18.6).  Hence, the supplementary ELG list
appears to be dominated by intermediate luminosity galaxies with a significantly
lower rate of star-formation activity (lower equivalent widths, redder colors)
than the ELGs in the main sample.

%********************************* REFERENCES***************************

%\clearpage

% Now comes the reference list.  In this document, we used \cite to call
% out citations, so we must use \bibitem in the reference list, which
% means we use the LaTeX thebibliography environment.  Please note that
% \begin{thebibliography} is followed by a null argument.  If you forget
% this, mayhem ensues, and LaTeX will say "Perhaps a missing item?" when
% you run it.  Do not call us, do not send mail when this happens.  Put
% the silly {} after the \begin{thebibliography}.
%
% Each reference has a \bibitem command to define the citation format
% to be placed in the text (in []) and the symbolic tag used for 
% cross referencing (in {}).
%
% See sample1.tex, or the AASTeX guide, for an alternative to the \cite-
% \bibitem command.

\clearpage
\renewcommand{\arraystretch}{.6}
\begin{deluxetable}{rrrllcccrrcl}
\scriptsize
%\rotate
\tablecolumns{12}
\tablewidth{0pt}
\tablecaption{List of Candidate ELGs\label{table:tab1}}
\tablenum{1}
\tablehead{
\colhead{KISSR}&\colhead{Field}&\colhead{ID}&\colhead{R.A.}&\colhead{Dec.}&\colhead{B}
&\colhead{B$-$V}&\colhead{z$_{KISS}$}&\colhead{Flux\tablenotemark{a}}
&\colhead{EW}&\colhead{Qual.}&\colhead{Comments}\\
\colhead{\#}&&&\colhead{(J2000)}&\colhead{(J2000)}&&&&&\colhead{[\AA]}&&\\
\colhead{(1)}&\colhead{(2)}&\colhead{(3)}&\colhead{(4)}
&\colhead{(5)}&\colhead{(6)}&\colhead{(7)}
&\colhead{(8)}&\colhead{(9)}&\colhead{(10)}&\colhead{(11)}&\colhead{(12)}
}
\startdata
    1&F1215 & 4723&12 15 06.9&29 01 10.0&  17.35&   0.53& 0.0266&  120&   53& 1&\phm{imaveryveryverylongstring}\\
    2&F1215 & 4362&12 15 23.9&29 10 46.8&  16.79&   0.68& 0.0300&  202&   27& 1&\phm{imaveryveryverylongstring}\\
    3&F1215 & 4188&12 15 36.9&28 59 29.4&  16.41&   0.81& 0.0283&  204&   23& 1&\phm{imaveryveryverylongstring}\\
    4&F1215 & 3026&12 16 46.4&29 26 10.5&  18.04&   0.78& 0.0619&  141&   44& 1&\phm{imaveryveryverylongstring}\\
    5&F1215 & 2774&12 17 10.1&28 57 53.1&  19.02&   0.52& 0.0590&   53&   46& 2&\phm{imaveryveryverylongstring}\\
    6&F1215 & 2005&12 17 56.9&29 07 16.6&  16.88&   0.63& 0.0510&  180&   23& 1&\phm{imaveryveryverylongstring}\\
    7&F1215 & 1851&12 18 11.9&28 43 39.0&  18.55&   0.72& 0.0837&   52&   40& 1&\phm{imaveryveryverylongstring}\\
    8&F1215 & 1748&12 18 12.2&29 15 06.3&  16.60&   0.60& 0.0269&  149&   28& 1& CG 167, Was 51                \\
    9&F1215 & 1639&12 18 19.3&29 15 13.3&  15.97&   0.88& 0.0492&  309&   23& 1&UGC 7342                       \\
   10&F1215 & 1628&12 18 20.2&29 14 48.2&  17.80&   0.82& 0.0696&  339&   71& 1&\phm{imaveryveryverylongstring}\\
 \\
   11&F1215 & 1631&12 18 23.4&28 58 10.7&  17.14&   0.65& 0.0779&  464&   90& 1&\phm{imaveryveryverylongstring}\\
   12&F1215 & 1627&12 18 23.5&28 58 29.4&  17.16&   0.45& 0.0767&  459&  110& 1&\phm{imaveryveryverylongstring}\\
   13&F1215 & 1485&12 18 33.1&28 56 33.1&  18.56&   0.62& 0.0763&   89&   73& 1&\phm{imaveryveryverylongstring}\\
   14&F1215 & 1187&12 18 41.8&29 41 27.1&  18.39&   0.79& 0.0784&   85&   54& 1&\phm{imaveryveryverylongstring}\\
   15&F1215 & 1040&12 18 58.6&29 09 03.4&  20.52&   0.84& 0.0582&   26&   58& 3&\phm{imaveryveryverylongstring}\\
   16&F1215 &  886&12 19 06.3&29 22 04.8&  18.62&   0.88& 0.0703&   54&   23& 2&\phm{imaveryveryverylongstring}\\
   17&F1215 &  978&12 19 06.8&28 45 39.8&  18.15&   0.93& 0.0760&  128&   50& 1&\phm{imaveryveryverylongstring}\\
   18&F1215 &  767&12 19 18.0&29 05 59.9&  18.50&   0.67& 0.0839&   75&   49& 1&\phm{imaveryveryverylongstring}\\
   19&F1215 &  745&12 19 18.7&29 10 34.0&  21.36&   0.72& 0.0577&   31&   39& 3&\phm{imaveryveryverylongstring}\\
   20&F1215 &  788&12 19 21.4&28 42 04.9&  19.66&   0.67& 0.0739&   42&   82& 2&\phm{imaveryveryverylongstring}\\
 \\
   21&F1215 &  430&12 19 34.1&29 47 53.1&  18.36&   0.64& 0.0671&  133&   59& 1&\phm{imaveryveryverylongstring}\\
   22&F1215 &  509&12 19 37.6&28 55 55.1&  19.68&   0.85& 0.0796&   32&   46& 3&\phm{imaveryveryverylongstring}\\
   23&F1215 &  504&12 19 39.1&28 51 40.9&  19.05&   1.14& 0.0831&   73&   45& 2&\phm{imaveryveryverylongstring}\\
   24&F1215 &  387&12 19 44.2&29 13 06.7&  18.39&   0.80& 0.0843&   91&   42& 1&\phm{imaveryveryverylongstring}\\
   25&F1215 &  361&12 19 49.2&28 58 24.8&  18.59&   1.09& 0.0825&   78&   33& 2&\phm{imaveryveryverylongstring}\\
   26&F1215 &  226&12 19 49.9&29 40 17.4&  19.99&   1.43& 0.0157&   96&   80& 3&\phm{imaveryveryverylongstring}\\
   27&F1215 &  228&12 19 50.6&29 36 52.3&  11.46&   0.99& 0.0092& 1254&    9& 3&UGC 7377                       \\
   28&F1220 & 3903&12 20 01.4&29 28 07.6&  18.43&   0.65& 0.0761&  103&   66& 1&\phm{imaveryveryverylongstring}\\
   29&F1220 & 3863&12 20 06.8&29 16 50.3&  11.53&   0.99& 0.0065& 1715&    3& 2&UGC 7386, CG 170               \\
   30&F1220 & 3801&12 20 16.2&28 51 19.4&  19.08&   0.98& 0.0799&   40&   52& 2&\phm{imaveryveryverylongstring}\\
\enddata
\tablenotetext{}{Note.--- The complete version of this table is presented in the
electronic edition of the Journal.  A portion is shown here for guidance regarding
its content and format.}
\tablenotetext{a}{Units of 10$^{-16}$ erg/s/cm$^2$}
\end{deluxetable}

\clearpage

\renewcommand{\arraystretch}{.6}

\begin{deluxetable}{rrrllcccrrcl}
\scriptsize
\tablecolumns{12}
\tablewidth{0pt}
\tablecaption{List of 4$\sigma$ Candidate ELGs\label{table:tab2}}
\tablenum{2}
\tablehead{
\colhead{KISSRx}&\colhead{Field}&\colhead{ID}&\colhead{R.A.}&\colhead{Dec.}&\colhead{B}
&\colhead{B$-$V}&\colhead{z$_{KISS}$}&\colhead{Flux\tablenotemark{a}}
&\colhead{EW}&\colhead{Qual.}&\colhead{Comments}\\
\colhead{\#}&&&\colhead{(J2000)}&\colhead{(J2000)}&&&&&\colhead{[\AA]}&&\\
\colhead{(1)}&\colhead{(2)}&\colhead{(3)}&\colhead{(4)}
&\colhead{(5)}&\colhead{(6)}&\colhead{(7)}
&\colhead{(8)}&\colhead{(9)}&\colhead{(10)}&\colhead{(11)}&\colhead{(12)}
}
\startdata
   1&F1215 & 4333&12 15 26.2&29 06 48.6&  17.74&   0.54& 0.0301&   65&   36& 1&\phm{imaveryveryverylongstring}\\
    2&F1220 & 2840&12 21 23.6&28 40 15.8&  18.64&   1.01& 0.0738&   39&   25& 2&\phm{imaveryveryverylongstring}\\
    3&F1220 &  945&12 23 49.3&29 43 04.1&  18.68&   0.57& 0.0645&   32&   30& 2&\phm{imaveryveryverylongstring}\\
    4&F1225 & 5035&12 25 19.0&28 50 12.2&  22.91&   2.48& 0.0017&   60&  129& 2&\phm{imaveryveryverylongstring}\\
    5&F1225 &   48&12 30 03.0&29 04 17.3&  18.72&   0.69& 0.0639&   47&   37& 2&\phm{imaveryveryverylongstring}\\
    6&F1235 & 1105&12 37 56.7&29 41 06.2&  19.74&   0.33& 0.0545&   19&   44& 3&\phm{imaveryveryverylongstring}\\
    7&F1240 & 4669&12 40 01.8&29 32 09.8&  21.08&   1.83& 0.0833&   37&   60& 2&\phm{imaveryveryverylongstring}\\
    8&F1245 & 1131&12 48 51.2&28 58 04.9&  17.33&   0.98& 0.0346&   30&    7& 1&\phm{imaveryveryverylongstring}\\
    9&F1250 & 2820&12 51 24.9&29 23 52.2&  18.39&   1.00& 0.0294&   46&   24& 2&\phm{imaveryveryverylongstring}\\
   10&F1250 & 2180&12 52 14.8&29 25 43.3&  16.33&   0.70& 0.0608&   69&    8& 2&\phm{imaveryveryverylongstring}\\
 \\
   11&F1255 & 1837&12 57 49.9&29 39 15.4&  15.25&   0.42& 0.0133&   57&    5& 1&UGC 8076                       \\
   12&F1300 & 2861&13 01 09.6&29 03 55.9&  18.59&   0.19& 0.0300&   27&   34& 2&\phm{imaveryveryverylongstring}\\
   13&F1300 & 2494&13 01 34.1&28 58 55.9&  18.58&   0.92& 0.0849&   48&   21& 2&\phm{imaveryveryverylongstring}\\
   14&F1300 &  675&13 03 47.4&29 42 39.7&  17.93&   0.51& 0.0866&   61&   37& 2&\phm{imaveryveryverylongstring}\\
   15&F1300 &  388&13 04 22.1&28 45 52.9&  18.58&   0.19& 0.0190&   34&   43& 2&\phm{imaveryveryverylongstring}\\
   16&F1305 & 5898&13 05 31.4&29 38 44.7&  19.95&   1.91& 0.0605&   54&   49& 2&\phm{imaveryveryverylongstring}\\
   17&F1310 &  151&13 14 52.8&28 43 21.9&  17.42&   0.37& 0.0763&   26&   14& 2&\phm{imaveryveryverylongstring}\\
   18&F1315 & 4390&13 15 06.4&29 36 58.7&  17.27&   1.16& 0.0623&  212&   29& 2&\phm{imaveryveryverylongstring}\\
   19&F1315 & 3896&13 15 50.6&28 37 16.8&  19.24&   0.84& 0.0203&   36&   27& 3&\phm{imaveryveryverylongstring}\\
   20&F1325 & 4575&13 24 54.2&28 49 49.8&  17.67&   0.25& 0.0559&   62&   37& 2&\phm{imaveryveryverylongstring}\\
 \\
   21&F1330 & 3167&13 32 24.8&29 05 04.7&  17.77&   0.75& 0.0447&   43&   12& 2&\phm{imaveryveryverylongstring}\\
   22&F1330 &  466&13 34 38.3&29 27 38.5&  18.63&   0.70& 0.0393&   21&   17& 2&\phm{imaveryveryverylongstring}\\
   23&F1335 & 3771&13 35 37.1&28 58 17.3&  18.98&   0.59& 0.0369&   38&   38& 2&\phm{imaveryveryverylongstring}\\
   24&F1335 & 1550&13 38 14.1&29 43 07.2&  18.69&   0.65& 0.0367&   25&   15& 2&\phm{imaveryveryverylongstring}\\
   25&F1340 & 3564&13 41 03.4&29 19 42.7&  18.55&   1.24& 0.0827&   16&    6& 3&\phm{imaveryveryverylongstring}\\
   26&F1345 & 4781&13 44 50.1&29 13 55.5&  17.69&   0.94& 0.0298&   35&    8& 2&\phm{imaveryveryverylongstring}\\
   27&F1345 & 2614&13 46 56.4&29 39 29.3&  17.46&   0.63& 0.0294&   87&   26& 2& Mk 274                        \\
   28&F1345 & 2220&13 47 25.2&29 37 31.9&  18.36&   0.64& 0.0775&   97&   57& 1&\phm{imaveryveryverylongstring}\\
   29&F1345 & 1931&13 47 54.7&28 41 06.5&  17.63&   1.12& 0.0277&   67&   10& 1&\phm{imaveryveryverylongstring}\\
   30&F1345 &  875&13 49 04.0&28 48 59.3&  17.97&   0.34& 0.0695&   29&   14& 2&\phm{imaveryveryverylongstring}\\
\enddata
\tablenotetext{}{Note.--- The complete version of this table is presented in the
electronic edition of the Journal.  A portion is shown here for guidance regarding
its content and format.}
\tablenotetext{a}{Units of 10$^{-16}$ erg/s/cm$^2$}
\end{deluxetable}

\end{document}